\documentclass[twocolumn]{aastex631}

\usepackage{CJKutf8}

\usepackage{float}
\usepackage{multirow}
\usepackage{subfigure}
\usepackage{booktabs}
\usepackage{amsmath}
\usepackage{graphicx}
\usepackage[section]{placeins}

\begin{document}


\title{Constraint on the physical origin of Gamma-Ray Burst prompt emission via its nondetected diffuse neutrino emission}

\correspondingauthor{Hou-Jun L\"{u} }
\email{lhj@gxu.edu.cn}

\author[0009-0007-5062-3221]{Yang-Dong-Jun Ou}
\affiliation{Guangxi Key Laboratory for Relativistic Astrophysics, School of Physical Science and Technology, Guangxi University, Nanning 530004, People's Republic of China; {\it lhj@gxu.edu.cn}}

\author[0000-0001-6396-9386]{Hou-Jun L\"{u} }
\affiliation{Guangxi Key Laboratory for Relativistic Astrophysics, School of Physical Science and Technology, Guangxi University, Nanning 530004, People's Republic of China; {\it lhj@gxu.edu.cn}}

\author[0000-0001-5681-6939]{Jia-Ming Chen }
\affiliation{Department of Astronomy, School of Physics and Astronomy, Yunnan University, Kunming, Yunnan 650091, People's Republic of China}

\author[0000-0003-1511-5567]{Ben-Yang Zhu}
\affiliation{Key Laboratory of Dark Matter and Space Astronomy, Purple Mountain Observatory, Chinese Academy of Sciences, 210033 Nanjing, Jiangsu, China }
\affiliation{School of Astronomy and Space Science, University of Science and Technology of China, 230026 Hefei, Anhui, China}

\author[0000-0002-7044-733X]{En-Wei Liang}
\affiliation{Guangxi Key Laboratory for Relativistic Astrophysics, School of Physical Science and Technology, Guangxi University, Nanning 530004, People's Republic of China; {\it lhj@gxu.edu.cn}}

\begin{abstract}
The physical origin of prompt emission in gamma-ray bursts (GRBs) remains an open question since it has been studied more than half a century. Three alternative models (i.e. dissipative photosphere, internal shock, and Internal-Collision-induced MAgnetic Reconnection and Turbulence, ICMART) have been proposed to interpret the observations of GRB prompt emission, but none of them can fully interpret all of the observational data collected so far. The question is what is the fraction of these three theoretical models in the prompt emission of GRBs. In this paper, we propose to utilize an innovative method and constrain the fraction of GRB prompt emission models via its nondetected diffuse neutrinos. By adopting two methods (e.g., summing up the individual GRB contributions and assumed luminosity functions of GRB) to calculate diffuse neutrino flux of GRBs for given the benchmark parameters of $\Gamma=300$ and $\varepsilon_{p} \text{/} \varepsilon_{e}=10$, both approaches indicate that most GRBs should be originated from the ICMART model. Moreover, we find that the fractions of the dissipative photosphere model, the internal shock model, and the ICMART model are constrained to be [0, $0.5\%$], [0, $1.1\%$], and [$98.9\%$, 1], respectively, for the method of summing up the individual GRB contributions. For the method of luminosity functions, the fractions of above three models are constrained to be [0, $6.1\%$], [0, $8.2\%$], and [$91.8\%$, 1], respectively. However, such fractions of different models are also dependent on the parameters of $\Gamma$ and $\varepsilon_{p} \text{/} \varepsilon_{e}$.

\end{abstract}


\begin{CJK*}{UTF8}{gbsn}
\section{Introduction}\label{sec:1}
\end{CJK*}
Despite the discovery of gamma-ray bursts (GRBs) more than half a century ago, the origin of the prompt emission in GRBs remains an unresolved question \citep{2018pgrb.book.....Z}. From the theoretical point of view, the observed prompt emission of GRBs can be
explained by the dissipative
photosphere model \citep{2005ApJ...628..847R}, internal shock model \citep{1994ApJ...430L..93R}, or the Internal-Collision-induced MAgnetic Reconnection and Turbulence (ICMART) model \citep{2011ApJ...726...90Z}. From the observational point of view, a purely quasi-thermal component is observed in GRB 090902B which is thought to be from the dissipative photosphere model \citep{2009ApJ...706L.138A,2010ApJ...709L.172R}. In contrast, a purely nonthermal emission is also observed in GRB 080916C which is consistent with the prediction of the ICMART model \citep{2009Sci...323.1688A,2009ApJ...700L..65Z,2011ApJ...726...90Z}. Moreover, there is a small fraction of GRBs whose observed spectrum is composed of both thermal and nonthermal components, such as GRBs 100724B, 110721A, 120323A, 160625B, 081221, and 211211A \citep{2011ApJ...727L..33G,2013ApJ...770...32G,2012ApJ...757L..31A,2017ApJ...849...71L,2018ApJ...866...13H,2023ApJ...943..146C}.
Therefore, it suggests that the jet composition of GRBs may be diverse and case by case, with different origins.

On the other hand, GRBs are the most energetic phenomena at cosmological distance and have been proposed to be the potential sources for extragalactic neutrinos \citep[e.g.,][]{1997PhRvL..78.2292W,2004APh....20..429G,2006PhRvD..73f3002M,2006PhRvL..97e1101M,2006ApJ...651L...5M,2009ApJ...691L..67W,2012PhRvL.108w1101H,2012JCAP...11..058G,2013PhRvL.110l1101Z,2016PhRvD..93l3004L,2017ApJ...848L...4K,2021JCAP...05..034P,2023ApJ...950..190M,2024ApJ...974..185M,2024JCAP...10..054D}. However, no robust associations of neutrinos with GRBs are found, and only the upper limits of nondetected neutrinos are reported in some GRBs \citep{2010ApJ...710..346A,2015ApJ...805L...5A,2016ApJ...824..115A,2017ApJ...843..112A,2022ApJ...939..116A,2024ApJ...964..126A}. Based on the results of nondetected neutrinos, significant constraints have been placed on the parameter space of the GRB jet compositions \citep[e.g.,][]{2012Natur.484..351I,2012ApJ...752...29H,2012PhRvD..85b7301L,2013ApJ...772L...4G,2015ApJ...805L...5A,2016ApJ...824..115A,2017ApJ...843..112A,2022ApJ...941L..10M,2023ApJ...944..115A,2024arXiv240816748V,2024ApJ...976..174O,2025ApJ...987...79L}. For example, the most significant constraint is from the brightest-of-all-time GRB 221009A, and it suggests that the dissipative photosphere model is highly disfavored, and the internal shock model can only survive in a small range of parameter space \citep[e.g.,][]{2022ApJ...941L..10M,2023ApJ...944..115A,2024arXiv240816748V}. 

Furthermore, the IceCube Collaboration claimed that the prompt emission of all observed GRBs cannot contribute more than 1$\%$ of the flux of diffuse neutrinos observed by IceCube \citep{2015ApJ...805L...5A,2017ApJ...843..112A,2022ApJ...939..116A}. If this is the case, by assuming that the prompt emission of all observed GRBs originates from a single model, the IceCube Collaboration adopted a set of 1172 GRBs to calculate the flux of diffuse neutrinos from the photosphere model, the internal shock model, as well as the ICMART model \citep{2017ApJ...843..112A}. It is found that the flux of diffuse neutrinos from both the photosphere model and the internal shock model is above the upper limit of IceCube. However, it does not mean that the dissipative photosphere model could be ruled out in all observed GRBs. For example, the prompt emission of GRBs may originate from different models (e.g., the photosphere model, the internal shock model, and the ICMART model) rather than from a single model, but we cannot confirm what proportion of those three models are.

Recently, systematic analyses of the time-resolved spectrum of Fermi GRBs have revealed that most GRB jets should contain a significant magnetic energy component. Specifically, \citet{2024ApJ...972....1L} found that 22. 7\% of GRBs that exhibit nonthermal spectra have $1+\sigma_{0} > 10$, indicating that their prompt emission is likely to originate from magnetized dissipation processes. On the other hand, previous studies have suggested that more than half of GRBs with prompt emission have photospheric emission \citep{2020ApJ...893..128A,2024ApJS..275....9C}. These results raise the interesting questions: is the GRB jet composition dominated by Poynting flux or matter? Is the fraction of GRBs whose prompt emission originates from the dissipative photosphere model and the ICMART model?

In this paper, we employ an independent method and adopt emission of diffuse neutrinos from GRBs to constrain the fraction of GRB prompt emission models. The method of calculation of the high-energy diffuse neutrino emission from GRBs is presented in Sec. \ref{sec:2}. In Sec. \ref{sec:3}, we show the constraints on the fraction of GRB prompt emission models via diffuse neutrino emission. The conclusions with some discussions are presented in Sec. \ref{sec:4}. Throughout this paper, we adopt the cosmological parameters with \(H_{0}=70 
\text{ km s}^{-1}\text{ Mpc}^{-1}\),  \(\Omega_{m}=0.3\), \(\Omega_{\Lambda}=0.7\), and the convention \(Q_{x} = Q/10^{x}\) in centimeter gram seconds.

\section{High-energy diffuse neutrino emission from GRBs}
\label{sec:2}

\subsection{Calculation of the diffuse neutrino emission from GRBs}
In this section, we adopt two methods to calculate diffuse neutrino flux of GRBs. One is following the procedure used by the IceCube Collaboration and ANTARES Collaboration \citep[e.g.,][]{2017ApJ...843..112A,2021MNRAS.500.5614A}, which involves a summing up the contributions of individual observed GRBs to the diffuse neutrino flux, and this method requires the use of real GRB samples. The other one is not to invoke the real GRB samples but employs an assumed GRB luminosity function \citep[e.g.,][]{2015JCAP...09..036T,2017ApJ...843...17X}.

By adopting the first approach mentioned above (e.g., summing up the contributions of individual observed GRBs to the diffuse neutrino flux), the quasi-diffuse neutrino flux of GRBs can be calculated by summing of the neutrino fluences from individual GRB and then rescaling the total fluence with the all-sky GRB event rate. In this work, we adopt the observed GRB rate of 667 $\rm yr^{-1}$, which is consistent with the value used in \citet{2017ApJ...843..112A} and \citet{2021MNRAS.500.5614A}. The method we use to calculate the neutrino spectrum for a single GRB was presented in \citet{2024ApJ...976..174O}. The quasi-diffuse neutrino flux is given by
\begin{equation}
     \begin{aligned}
    E_{\nu }^{2} \phi _{\nu} =\sum_{i=1}^{N_{\rm GRB}} (E_{\nu }^{2} \phi _{\nu})_{i}\frac{1}{4\pi }\frac{667}{N_{\rm GRB}}  \rm yr^{-1}, 
     \end{aligned}
 \end{equation}
where $N_{\rm GRB}$ is the number of GRBs in the selected sample. 

By adopting the second approach mentioned above (e.g., applying the assumed GRB luminosity function to the diffuse neutrino flux calculation), it is necessary to invoke an assumptions of regarding the event rate and luminosity function \citep[e.g.,][]{2010MNRAS.406.1944W,2015MNRAS.448.3026W,2015ApJ...812...33S}. Here, we use the model derived by \citet{2010MNRAS.406.1944W}, and the event rate is given by
\begin{equation}
     \begin{aligned}
\rho(z)=\rho_{0}\begin{cases}
(1+z)^{a} &  z<z_{\star} \\
\left(1+z_{\star}\right)^{a-b}(1+z)^{b} &  z \geq z_{\star}.
\end{cases}
     \end{aligned}
 \end{equation}
Here, \(\rho_{0}=1.3 \rm ~Gpc^{-3} yr^{-1}\), \(z_{\star}=3.1\), \(a=2.1\), and \(b=-1.4\). The luminosity function is given by
\begin{equation}
     \begin{aligned}
\phi (L)\propto \begin{cases}
(\frac{L }{L_{b}} )^{-\alpha_{\phi} } &  L<L_{b}, \\
(\frac{L }{L_{b}})^{-\beta_{\phi} } &  L \geq L_{b},
\end{cases}
     \end{aligned}
 \end{equation}
where \(\alpha_{\phi}=1.2\) and \(\beta_{\phi}=2.4\) are the indices in luminosity function and \(L_{b}=10^{52.5}\text { erg }\rm s^{-1}\). The quasi-diffuse neutrino flux of GRBs can be calculated as
\begin{equation}
     \begin{aligned}
\phi _{\nu}\left ( E_{\nu } \right )&=\frac{c}{4\pi H_{0}}\int_{0}^{z_{\max}}dz\int_{L_{\min}}^{L_{\max}}dL
\\&\times \frac{1}{(1+z)^{2}\sqrt{\Omega_{M}(1+z)^{3}+\Omega_{\Lambda}}}\\&\times 
\rho(z)\phi (L)\frac{dN_{\nu } \left ( E_{\nu } \right )}{dE_{\nu }}.
     \end{aligned}
 \end{equation}
Here, we adopt \(z_{\max}=10, L_{\min}=10^{49}\text { erg/s}\), and \(L_{\max}=10^{54}\text { erg/s}\) to do the calculations. It is quite different from the first method, which uses real photon spectral parameters for each GRB; we have to assume a constant of photon spectral parameters with low-energy index (\(\alpha = -1\)), high-energy index (\(\beta = -2\)), break energy  \(\varepsilon _{\gamma,b}=\frac{(\alpha-\beta)E_{\text{peak}}}{2+\alpha} = 0.3\text { MeV} \), and $E_{\rm peak}$ being the peak energy of spectra. One needs to note that in the luminosity function, \(L\) refers to the peak luminosity, whereas the neutrino flux calculation requires the average luminosity. Therefore, we use the Amati relation \citep{2006MNRAS.372..233A} and Yonetoku relation \citep{2004ApJ...609..935Y} to estimate isotropic energy and assume that the average luminosity is the isotropic energy divided by 10 s. The Amati relation and Yonetoku relation are expressed as
\begin{equation}
     \begin{aligned}
 \frac{E_{\rm peak,z}}{100\mathrm{~keV}}\simeq\frac{4}{5}\left(\frac{E_{\rm iso}}{10^{52}\mathrm{~erg}}\right)^{0.4},
     \end{aligned}
 \end{equation}
\begin{equation}
     \begin{aligned}
\frac{E_{\rm peak,z}}{100\mathrm{~keV}}\simeq1.8\left(\frac{L}{10^{52}\mathrm{~erg~s^{-1}}}\right)^{0.52}.
     \end{aligned}
 \end{equation}
 
Previous estimates of diffuse neutrino flux from GRBs are based on the assumption of the prompt emission of all observed GRBs originating from a single model. In this work, we simultaneously take into account the three possible different origins of GRB prompt emission models (e.g., the photosphere model, the internal shock model, and the ICMART model).

If we consider different models of prompt emission, the neutrino fluences may be different. This is because the radius of prompt emission GRBs is dependent on the selected model, and a larger radius of prompt emission can lead to a smaller value of \(p\gamma \) optical depth. The prompt emission of the dissipative photosphere model comes from the Thomson scattering photosphere, and the radius of the dissipative photosphere model is given by
\begin{equation}
     \begin{aligned}
R_{\rm ph} =4.33 \times 10^{12}~{\rm cm}~ L_{{\rm iso},52} \epsilon_{e,-1}^{-1}(\frac{\Gamma}{300})^{-3}, 
     \end{aligned}
 \end{equation}
where $L_{{\rm iso}}$ is the isotropic luminosity, \(\varepsilon _{e}\) is the fraction of jet dissipated energy going into the electrons, and \(\Gamma\) is bulk motion Lorentz factor. The radius of the internal shock model is given by
\begin{equation}
     \begin{aligned}
R_{\rm IS} = 5.4 \times 10^{14}~{\rm cm}~ (\frac{\Gamma}{300})^2 (\frac{\delta t_{\rm min}}{0.1~{\rm s}} ) (1+z)^{-1},
     \end{aligned}
 \end{equation}
where $\delta t_{\rm min}$ is the minimum variability time scale. In our calculations, we adopt $\delta t_{\rm min} = 0.01\rm s$. 
The ICMART model invokes the magnetized dissipation at a large radius $R_{\rm ICMART}\sim 1 \times 10^{15}~{\rm cm}$.

By adopting the radius from a specific model of prompt emission, one can calculate the flux of a diffuse neutrino based on the assumption that all GRBs originate from a single model. We use the symbols $E_{\nu }^{2} \phi _{\nu}^{\rm ph}$, $E_{\nu }^{2} \phi _{\nu}^{\rm IS}$ and $E_{\nu }^{2} \phi _{\nu}^{\rm ICMART}$ to donate the flux of diffuse neutrino from the dissipative photosphere model, the internal shock model, and the ICMART model, respectively. By considering the contributions from different origins of GRB model, the flux of quasi-diffuse neutrino is expressed as 
\begin{equation}
     \begin{aligned}
    E_{\nu }^{2} \phi _{\nu} &= E_{\nu }^{2} \phi _{\nu}^{\rm ph}\times  F_{\rm ph}+E_{\nu }^{2} \phi _{\nu}^{\rm IS}\times  F_{\rm IS}\\&+E_{\nu }^{2} \phi _{\nu}^{\rm ICMART}\times  F_{\rm ICMART},
     \end{aligned}
 \end{equation}
where $F_{\rm ph}$, $F_{\rm IS}$ and $F_{\rm ICMART}$ are the fraction of the dissipative photosphere model, the internal shock model, and the ICMART model, respectively. 

\subsection{GRB sample selected}
In order to calculate the flux of diffuse neutrino from GRBs, one needs to obtain the spectral properties of all observed GRBs. In our calculations, we collect a larger GRB samples, which includes 3739 GRB detected by the Fermi Gamma-ray Burst Monitor from July 2008 to June 2024, and the spectral properties of our adopted GRB sample are taken from \citet{2025ApJS..276...62C}. \citet{2025ApJS..276...62C} adopted the Band function to fit the spectra of observed GRBs, and it can be described as
{\tiny
\begin{equation}
     \begin{aligned}
N(E)=A\begin{cases}
(\frac{E}{100~\mathrm{keV}})^{\alpha }\mathrm{exp}\left[-\frac{(\alpha+2)E}{E_{\text{peak}}} \right ], E<\frac{(\alpha-\beta)E_{\text{peak}}}{2+\alpha}, \\
(\frac{E}{100~\mathrm{keV}})^{\beta }\mathrm{exp}(\beta -\alpha
)[\frac{(\alpha-\beta)E_{\text{peak}}}{(2+\alpha)100~\mathrm{keV}}]^{\alpha-\beta }, E\geq \frac{(\alpha-\beta)E_{\text{peak}}}{2+\alpha}
\end{cases},
     \end{aligned}
 \end{equation}}
where $A$ is the normalization of the spectral fit, $E_{\text{peak}}$ is peak energy, and $\alpha$ and $\beta$ are the low-energy and high-energy photon spectral indices, respectively. In our sample, there are only 203 GRBs that have redshift measurement, and we adopt $z=2.15$ for long GRBs and $z=0.5$ for short GRBs to do the calculations for those without redshift measurement \citep[e.g.,][]{2017ApJ...843..112A}. 

On the other hand, one needs to calculate the neutrino spectrum of an individual GRB by adopting its isotropic luminosity $L_{\rm iso}$ and the isotropic energy $E_{\rm iso}$. The isotropic luminosity can be calculated by
\begin{equation}
     \begin{aligned}
L_{\rm iso} =4\pi d_{L}^{2}kF_{\gamma}, 
     \end{aligned}
 \end{equation}
where $F_{\gamma}$ is the photon flux. $d_{L}$ and $k$ are the luminosity distance and the $k$-correction factor which can be defined by
\begin{equation}
     \begin{aligned}
d_{L}= \frac{c(1+z)}{H_{0}} \int_{0}^{z} \frac{\mathrm{d}z}{\sqrt{(1+z)^{3}\Omega_{m}+\Omega_{\Lambda}} } , 
     \end{aligned}
 \end{equation}
 \begin{equation}
     \begin{aligned}
k=\frac{\int_{1~\rm keV/(1+z)}^{10^{4}~\rm keV/(1+z)} EN(E)\mathrm{d}E }{\int_{e_{1}}^{e_{2}} EN(E)\mathrm{d}E }, 
     \end{aligned}
 \end{equation}
where $c$ is the speed of light and $e_{\rm 1}=8  ~\rm keV$ and $e_{\rm 2}=40,000~\rm keV$ are the minimum energy and maximum energy of the energy band of the detector, respectively.
The isotropic energy can be calculated by
\begin{equation}
     \begin{aligned}
E_{\rm iso} =\frac{4\pi d_{L}^{2}kS_{\gamma}}{(1+z)},  
     \end{aligned}
 \end{equation}
where $S_{\gamma}=T_{90}F_{\gamma}$ is the photon fluence and $T_{90}$ is the duration of a burst with the time interval between the epochs when 5\% and 95\% of the total fluence are collected by the detector.

\section{Constraints on the fraction of GRB prompt emission models via diffuse neutrino emission} \label{sec:3}

Since \citet{2017ApJ...843..112A} used 7 yr of data, which include 1172 GRBs, to derive an upper limit of the diffuse neutrino flux, and the data sample adopted in this work is much larger than that of previous studies, we utilize the latest publicly available point-source dataset \footnote{https://icecube.wisc.edu/data-releases/2021/01/all-sky-point-source-icecube-data-years-2008-2018/} covering 10 yr of data to perform an unbinned maximum likelihood analysis and obtain the upper limit of diffuse neutrino flux. Our analysis includes 2338 GRBs, and we assume a single power-law neutrino spectrum of the form \(E_{\nu }^{-2}\) over an energy range from 10 TeV to 100 PeV for each GRB. The derived upper limit of the diffuse neutrino flux is \(3.71 \times 10^{-11} \rm~GeV~cm^{-2}~sr^{-1}~s^{-1}\).

\begin{figure}[H]
      \centering
      \includegraphics[width=0.495\linewidth]{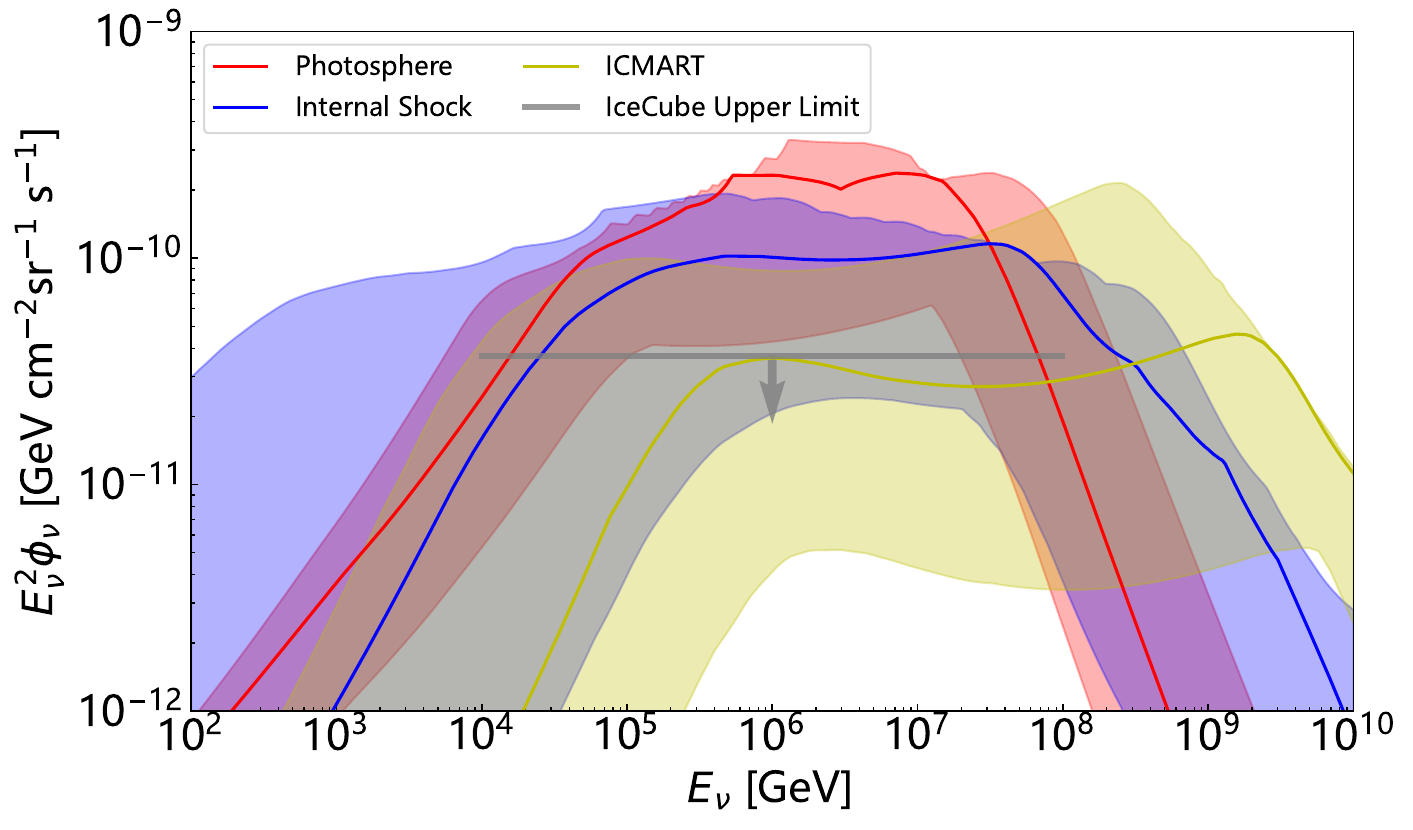}
      \includegraphics[width=0.495\linewidth]{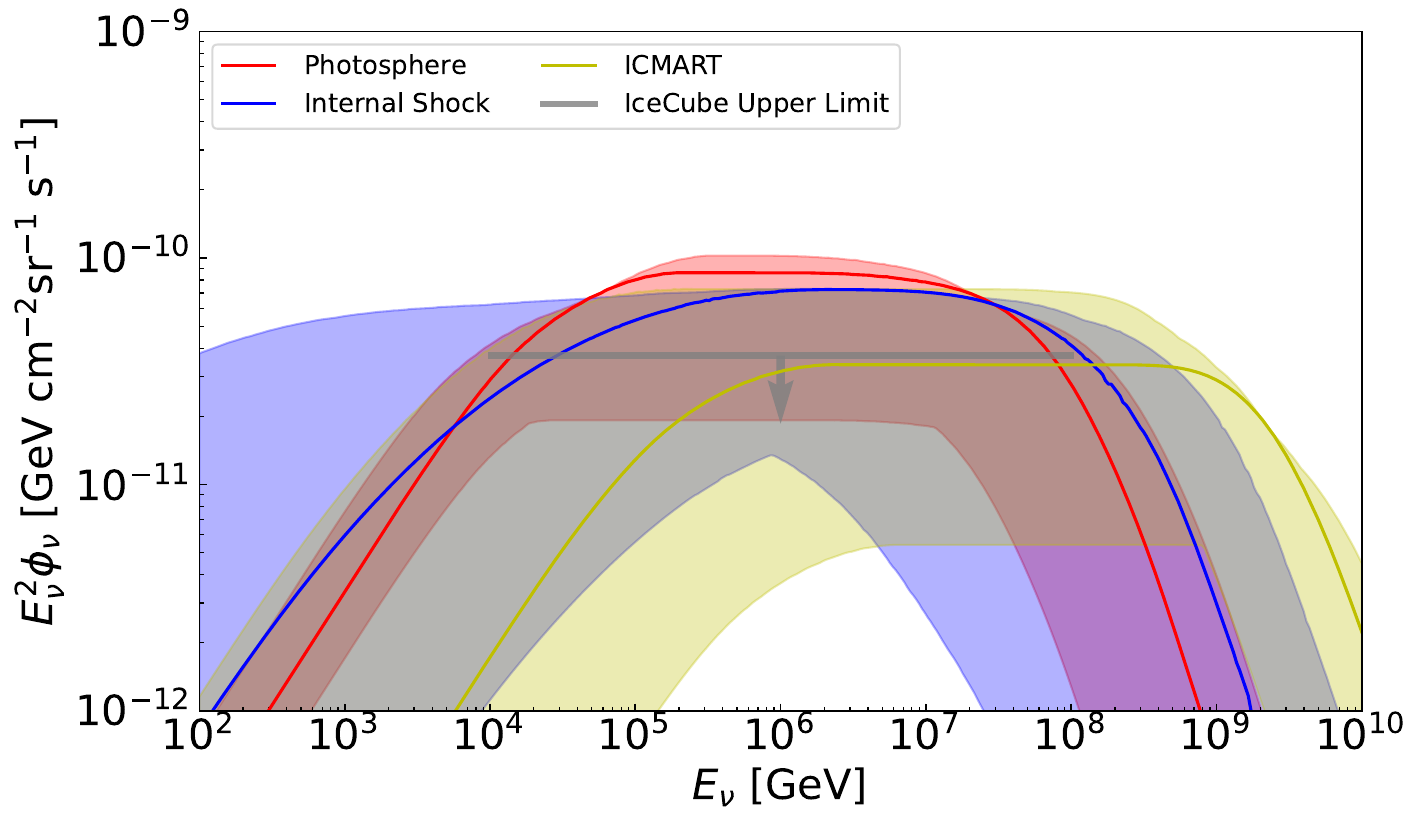}

      \caption{Predicted flux of diffuse neutrino for the benchmark parameters of $\varepsilon_{p} \text{/} \varepsilon_{e} = 10$ and $\Gamma = 300$ by assuming that all GRBs are originated from the signal model (e.g., the dissipative photosphere model, the internal shock model, and the ICMART model). The bands for the diffuse neutrino flux are displayed for $\varepsilon_{p} \text{/} \varepsilon_{e} = 3-10$ and $\Gamma = 100-500$. The gray line with arrows is the upper limit flux of diffuse neutrino. Left: results using the method of summing up the individual GRB contributions. Right: results using the method of assumed luminosity functions of the GRB.}
      \label{Fig:1}
  \end{figure}

In order to calculate the neutrino spectrum of an individual GRB, one needs to understand the unknown parameters of $\varepsilon_{p} \text{/} \varepsilon_{e}$ and $\Gamma$. By assuming that all GRBs originate from the signal model (e.g., the dissipative photosphere model, the internal shock model, and the ICMART model), Figure \ref{Fig:1} shows the predicted flux of the diffuse neutrino by adopting the range of parameters for $\varepsilon_{p} \text{/} \varepsilon_{e} = (3-10)$ and $\Gamma = (100-500)$ \citep{2017ApJ...843..112A}. Here, we adopt two methods to calculate the flux of diffuse neutrino, i.e., summing up the individual GRB contributions (left in Figure \ref{Fig:1}) and assumed luminosity functions of GRB (right in Figure \ref{Fig:1}). It is found that the predicted neutrino fluxes from both the photosphere and internal shock models exceed the upper limit of IceCube by invoking both methods, while the flux obtained from the assumed luminosity functions of GRB is slightly lower than that from the summing up the individual GRB contributions. It suggests that the prompt emission of all GRBs is not likely to originate from the signal model but rather from diverse models.

\begin{figure}[H]
      \centering
      \includegraphics[width=0.49\linewidth]{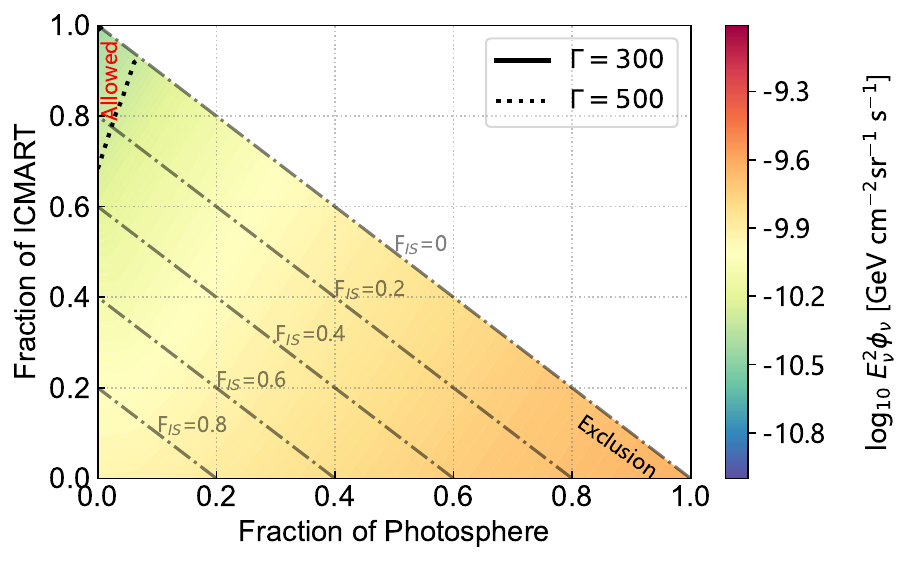}
      \includegraphics[width=0.49\linewidth]{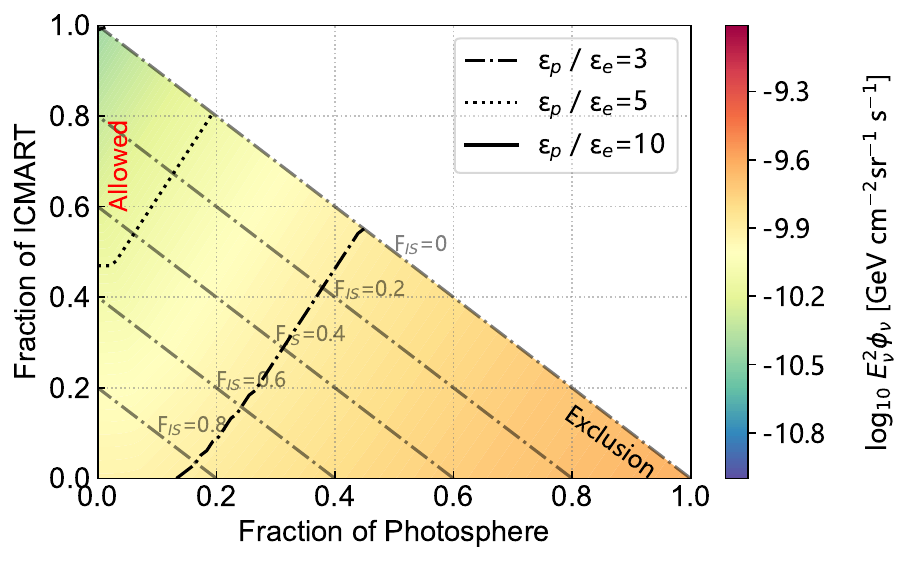}
      \caption{Constraint on the fraction of GRB prompt emission models by adopting the method of summing up the individual GRB contributions. Left: the constraint results for the varying value of $\Gamma=$ 300 and 500 with fixed $\varepsilon _{p } \text{/} \varepsilon _{e } = 10$. Right: the constraint results for the varying value of $\varepsilon _{p } \text{/} \varepsilon _{e }=3$, 5, and 10 with fixed $\Gamma = 300$.}
      \label{Fig:2}
  \end{figure}

\begin{figure}[H]
      \centering
      \includegraphics[width=0.49\linewidth]{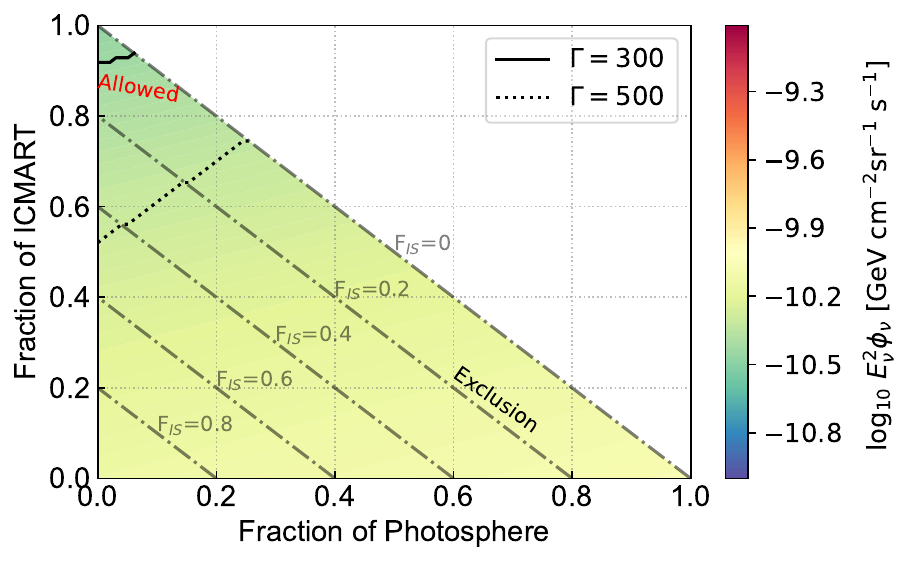}
      \includegraphics[width=0.49\linewidth]{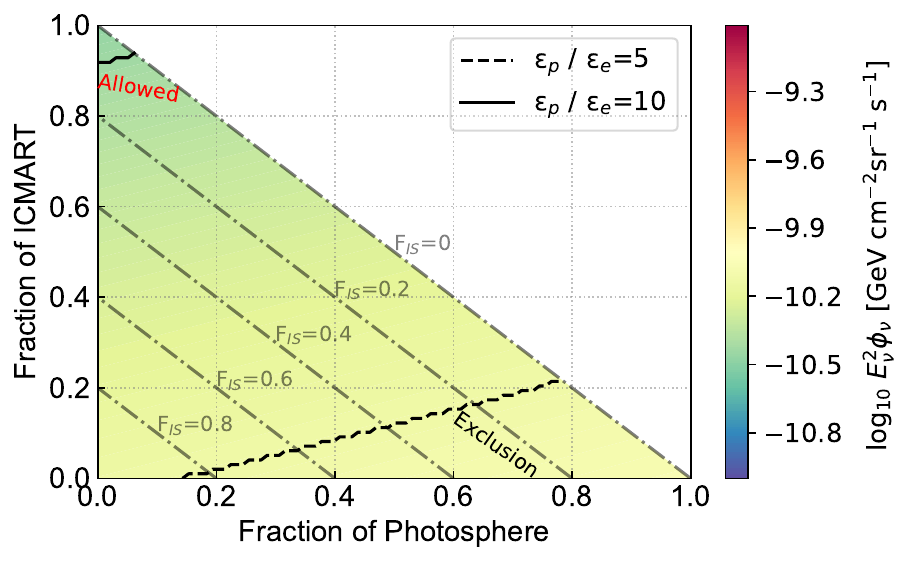}
      \caption{Constraint on the fraction of GRB prompt emission models by adopting the method of assumed luminosity functions of GRB. Left: the constraint results for the varying value of $\Gamma=$ 300 and 500 with fixed $\varepsilon _{p } \text{/} \varepsilon _{e } = 10$. Right: the constraint results for the varying value of $\varepsilon _{p } \text{/} \varepsilon _{e }=$ 5 and 10 with fixed $\Gamma = 300$.}
      \label{Fig:3}
  \end{figure}

In order to constrain the fraction of GRB prompt emission models, we adopt the varying parameters of $\varepsilon_{p} \text{/} \varepsilon_{e}=3$, 5, and 10 and $\Gamma=100$, 300, and 500 to do the calculations in Figure \ref{Fig:2} (the method of summing up the individual GRB contributions) and Figure \ref{Fig:3} (the method of assumed luminosity functions of GRB). We find that the case of $\Gamma=100$ and $\varepsilon_{p} \text{/} \varepsilon_{e}=10$ can be ruled out in both approaches. Moreover, no constraint can be placed for the case of $\Gamma=300$ and $\varepsilon_{p} \text{/} \varepsilon_{e}=3$ within the method of assumed luminosity functions of GRBs. By adopting the method of summing up the individual GRB contributions, taking into account the three possible models simultaneously, and fixed in $\varepsilon_{p} \text{/} \varepsilon_{e}=10$ and $\Gamma=300$, we find that the fraction ranges for the dissipative photosphere model, the internal shock model, and the ICMART model are constrained to be [0, $0.5\%$], [0, $1.1\%$], and [$98.9\%$, 1], respectively (see the solid line in Figure \ref{Fig:2}). For the method of assumed luminosity functions of GRBs, the fractions of the above three models are constrained to be [0, 6.1 \% ], [0, 8.2 \% ], and [ 91.8 \% , 1], respectively (see the solid line in Figure \ref{Fig:3}). However, the fraction ranges of these three models depend on the varying parameters of $\varepsilon_{p} \text{/} \varepsilon_{e}$ and $\Gamma$. For example, we fix $\varepsilon_{p} \text{/} \varepsilon_{e}=10$ and adopt the varying $\Gamma=100$, 300, and 500 and fix $\Gamma=300$ and adopt the varying $\varepsilon_{p} \text{/} \varepsilon_{e}=3$, 5, and 10. We find that the fraction ranges of these three models also change with both $\Gamma$ and $\varepsilon_{p} \text{/} \varepsilon_{e}$ (see Figure \ref{Fig:2} and \ref{Fig:3}, and Table \ref{table1} and \ref{table2}).

The above results suggests that the dissipative photosphere model and the internal shock model still survive in the prompt emission of a small faction of GRBs for the benchmark parameters of $\Gamma$ and $\varepsilon_{p} \text{/} \varepsilon_{e}$, and most of GRBs should originate from the ICMART model to avoid the flux of diffuse neutrinos above the upper limit of IceCube.

\section{Conclusion and discussion}\label{sec:4}
The multimessenger astronomy approach is a powerful tool to help us understand the physical origin of astronomical sources. GRBs are the most luminous explosions in the Universe and have rapidly advanced in recent years \citep{2018pgrb.book.....Z}. Especially, the jet composition of GRBs remains an open question \citep{2011CRPhy..12..206Z}. The direct method to identify the jet composition of GRBs is to use the types of observed spectra for GRB prompt emission, but the uncertainty of the measurements is still very large \citep{2011ApJ...730..141Z,2015ApJ...801..103G,2024ApJ...972....1L,2024ApJS..275....9C}. In this paper, we propose an independent method to constrain the fraction of GRB prompt emission models based on the nondetected high-energy diffuse neutrino, which is possibly associated with GRBs. For given the benchmark parameters of GRBs, i.e., $\Gamma=300$ and $\varepsilon_{p} \text{/} \varepsilon_{e}=10$, by adopting the method of summing up the individual GRB contributions, we find that the fractions of the dissipative photosphere model, the internal shock model, and the ICMART model are range in  [0, $0.5\%$], [0, $1.1\%$], and [$98.9\%$, 1], respectively. By adopting the method of assumed luminosity functions of GRBs, the corresponding fractions of above three models are [0, $6.1\%$], [0, $8.2\%$], and [$91.8\%$, 1], respectively. However, such fractions of different models are also dependent on the parameters of $\Gamma$ and $\varepsilon_{p} \text{/} \varepsilon_{e}$ (see Figure \ref{Fig:2} and \ref{Fig:3} and Table \ref{table1} and \ref{table2}).

\cite{2024ApJ...972....1L} analyzed the time-resolved spectrum for all Fermi GRBs with known redshift to estimate the magnetization factor ($\sigma_{0}$) and found that the lower limits of $1+\sigma_{0}$ for $22.7\%$ of the nonthermal spectrum (225 out of 318) are larger than 10. By adopting the benchmark parameters of $\Gamma=300$ and $\varepsilon_{p} \text{/} \varepsilon_{e}=10$ in our calculations, we estimate the lower limit of the fraction of the ICMART model and conclude that most of GRBs should originate from the ICMART model. Our results are not contradictory to those from \cite{2024ApJ...972....1L}, who obtained the lower limit of the magnetization factor. This suggests that most GRB jets should contain a significant magnetic-energy component, namely, magnetic dissipation. On the other hand, \cite{2017ApJ...843..112A} suggested that the internal shock model and the photospheric model are excluded at the $99\%$ confidence level for benchmark parameters of $\Gamma$ and $\varepsilon_{p} \text{/} \varepsilon_{e}$ when they assumed that the prompt emission of all observed GRBs originates from a single mode. However, by taking into account a certain fraction of the GRB model that is from the ICMART model, it is found that both the internal shock model and the photospheric model can survive for the benchmark parameters.

\begin{figure}[H]
      \centering
      \includegraphics[width=0.32\linewidth]{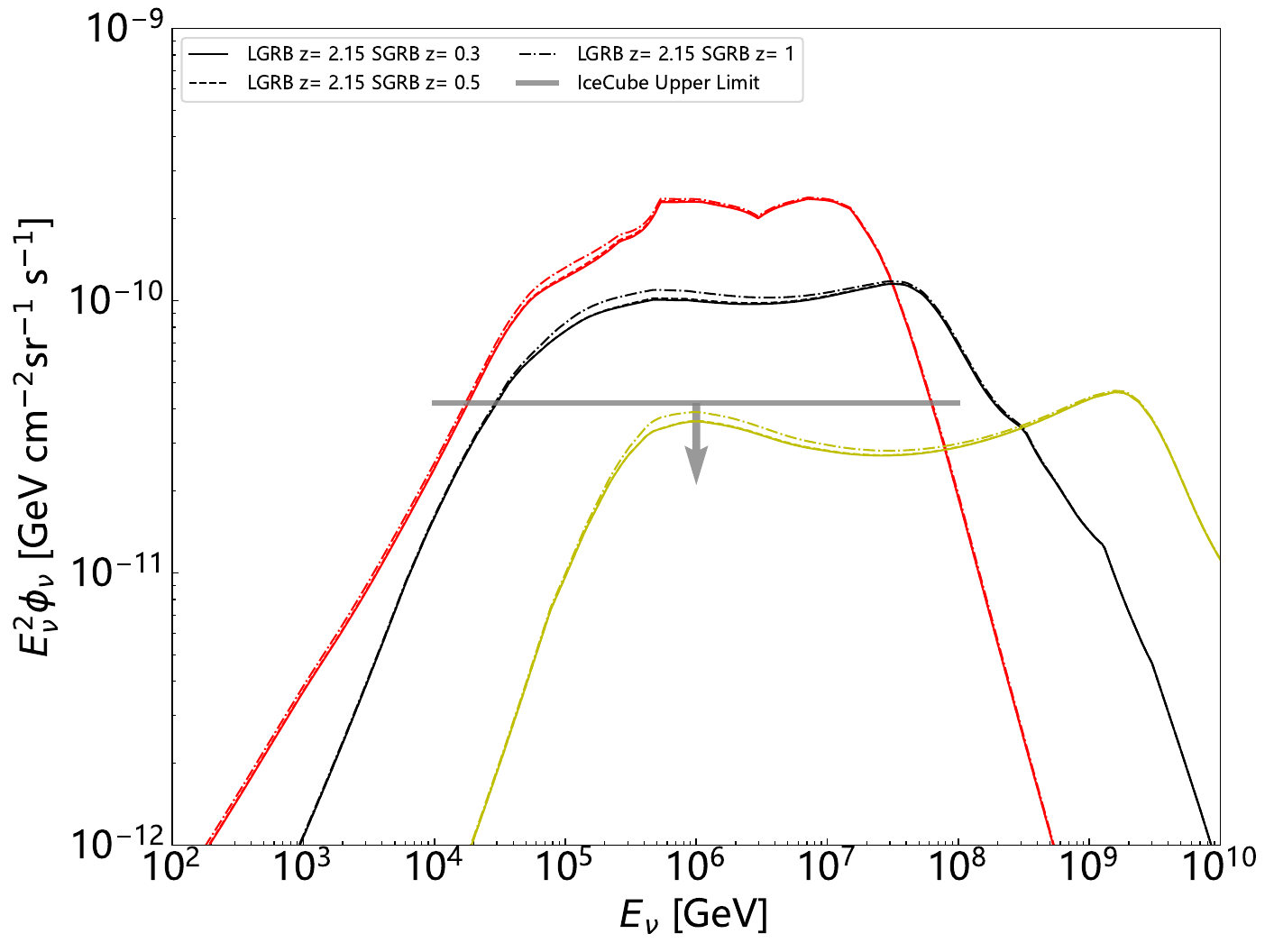}
      \includegraphics[width=0.32\linewidth]{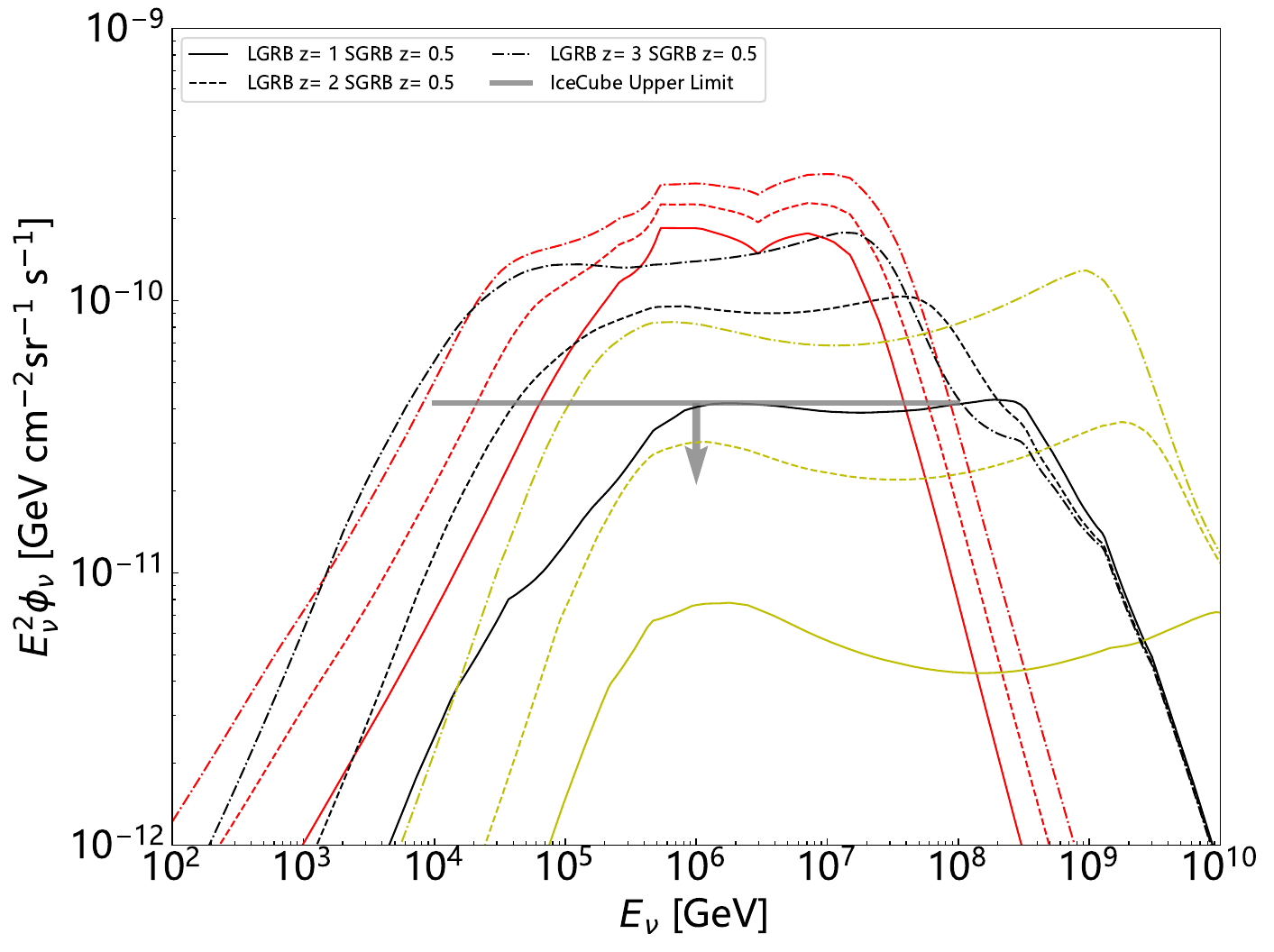}
      \includegraphics[width=0.32\linewidth]{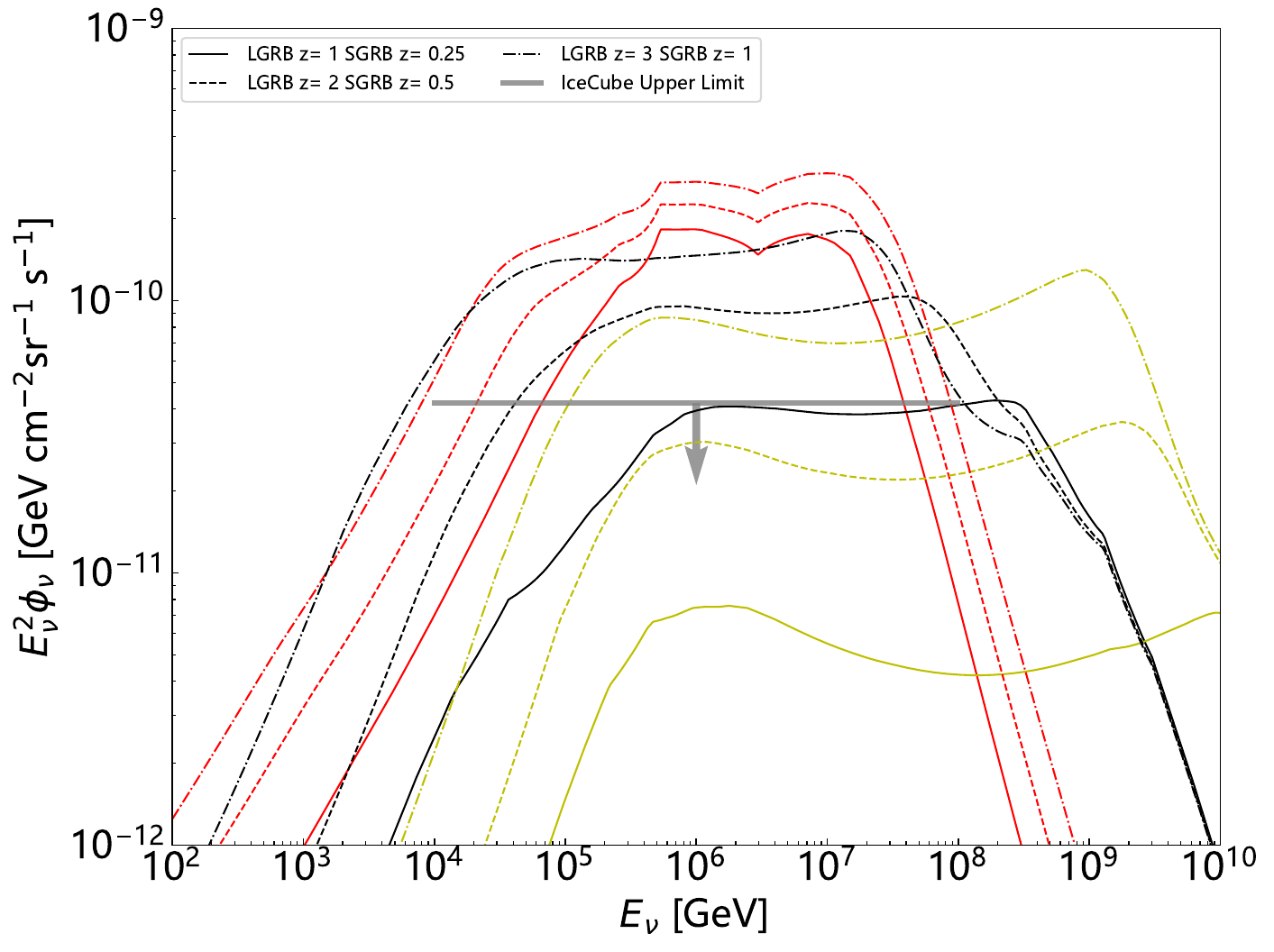}

      \caption{Predicted flux of diffuse neutrino for the benchmark parameters of $\varepsilon_{p} \text{/} \varepsilon_{e} = 10$ and $\Gamma = 300$ by assuming that all GRBs are originated from signal model with different mean redshift. Left: changed the mean redshift of short GRBs as $z=0.3$, 0.5, and 1.0 for fixed $z=2.15$ of long GRBs. Middle: changed the mean redshift of long GRBs as $z=1.0$, 2.0, and 3.0 for fixed $z=0.5$ of short GRBs. Right: simultaneously increasing the redshift of long GRBs ($z=1.0$, 2.0, and 3.0) and short GRBs ($z=0.25$, 0.5, and 1.0).}
      \label{Fig:4}
  \end{figure}

Our analysis also poses a curious question. To calculate the diffuse neutrino flux by summing up individual GRB contributions, we have to adopt average redshift values for the majority of GRBs that lack redshift measurements. This approach of using average values is also employed by the IceCube collaboration \citep{2017ApJ...843..112A}. In order to test the effect of varying redshift, we also adopt different values of redshift, and the predicted flux of diffuse neutrino for the benchmark parameters based on different redshift value assumption is shown in Figure \ref{Fig:4}. In the left panel of Figure \ref{Fig:4}, we test the effect of changing mean redshift of short GRBs $z=0.3$, 0.5, and 1.0 for fixed $z=2.15$ of long GRBs. The middle panel shows the effect of changing the mean redshift of long GRBs $z=1.0$, 2.0, and 3.0 for fixed $z=0.5$ of short GRBs, and the right panel in Figure \ref{Fig:4} shows the result of simultaneously increasing the redshifts of both long and short GRBs. We find that changing the mean redshift of short GRBs has a little impact on the results. That is because the fraction of short GRBs in our sample is not large enough, and the isotropic energy of short GRBs is usually lower than that of long GRBs. It means that the ability to produce neutrinos is also weaker by comparing the long GRBs. Moreover, it is found that the flux of diffuse neutrino is increasing with the increasing redshift. The reason is possibly that increasing redshift can also result in increasing $L_{\rm iso}$ for a given $F_{\gamma}$. In any case, increased $L_{\rm iso}$ could lead to an increase of \(p\gamma \) optical depth and an increase of the pion production efficiency if it is not saturated. We also find that the effect of diffuse neutrino flux with varying redshifts of GRBs is significantly larger for the large dissipation radius model (i.e. ICMART model). That is because the pion production efficiency is already saturated in most GRBs in a smaller dissipation radius model (i.e., the dissipative photosphere model), and further increases in \(p\gamma \) optical depth do not increase pion production efficiency. In the future, with the increasing number of GRB redshift measurements, the constraint on the fraction of GRB prompt emission models from the effect of redshift could be ignored.

On the other hand, it is worth noting that the constraints of this work strongly depend on the dissipation radius of photosphere, internal shock, and the ICMART models we adopted. In this work, we only adopt the typical values of the dissipation radius to constrain the physical origin of the GRB prompt emission \citep{2018pgrb.book.....Z}, and more precise constraints will be obtained once the dissipation radius can be accurately determined.

\section*{Acknowledgments}
We thank Xue-Zhao Chang and Si-Yuan Zhu for the helpful discussion. This work is supported by the Guangxi Science Foundation of the National (grant No. 2023GXNSFDA026007), the Natural Science Foundation of China (grant Nos. 12494574, 11922301 and 12133003), the Program of Bagui Scholars Program (LHJ), and the Guangxi Talent Program (“Highland of Innovation Talents”).

\bibliography{ms}
\bibliographystyle{aasjournal}

\begin{table}[H]
\centering
\caption{The fraction range of the photosphere model, the internal shock model, and the ICMART model within different $\varepsilon_{p} \text{/} \varepsilon_{e}$ and $\Gamma$ by adopting the method of summing up the individual GRB contributions.}

\begin{tabular}{|c|c|ccc|}
\hline
\multicolumn{1}{|l|}{\multirow{2}{*}{$\varepsilon_{p} \text{/} \varepsilon_{e}$}} & \multicolumn{1}{l|}{\multirow{2}{*}{$\Gamma$}} & \multicolumn{3}{c|}{Range of Fraction}                                                        \\ \cline{3-5} 
\multicolumn{1}{|l|}{}                                                            & \multicolumn{1}{l|}{}                          & \multicolumn{1}{c|}{Photosphere}     & \multicolumn{1}{c|}{Internal Shock}  & ICMART          \\ \hline
\multirow{3}{*}{10}                                                               & 100                                            & \multicolumn{1}{c|}{Rule out}        & \multicolumn{1}{c|}{Rule out}        & Rule out        \\ \cline{2-5} 
                                                                                  & 300                                            & \multicolumn{1}{c|}{{[}0, 0.5\%{]}} & \multicolumn{1}{c|}{{[}0, 1.1\%{]}} & {[}98.9\%, 1{]} \\ \cline{2-5} 
                                                                                  & 500                                            & \multicolumn{1}{c|}{{[}0, 7.1\%{]}}  & \multicolumn{1}{c|}{{[}0, 31.7\%{]}}    & {[}68.3\%, 1{]}   \\ \hline
3                                                                                 & \multirow{3}{*}{300}                           & \multicolumn{1}{c|}{{[}0, 44.9\%{]}} & \multicolumn{1}{c|}{No constraint}   & No constraint   \\ \cline{1-1} \cline{3-5} 
5                                                                                 &                                                & \multicolumn{1}{c|}{{[}0, 19.3\%{]}} & \multicolumn{1}{c|}{{[}0, 53.1\%{]}}   & {[}46.9\%, 1{]}   \\ \cline{1-1} \cline{3-5} 
10                                                                                &                                                & \multicolumn{1}{c|}{{[}0, 0.5\%{]}} & \multicolumn{1}{c|}{{[}0, 1.1\%{]}} & {[}98.9\%, 1{]} \\ \hline
\end{tabular}
\label{table1}
\end{table}

\begin{table}[H]
\centering
\caption{The fraction range of the photosphere model, the internal shock model, and the ICMART model within different $\varepsilon_{p} \text{/} \varepsilon_{e}$ and $\Gamma$ by adopting the method of assumed luminosity functions of GRB.}

\begin{tabular}{|c|c|ccc|}
\hline
\multicolumn{1}{|l|}{\multirow{2}{*}{$\varepsilon_{p} \text{/} \varepsilon_{e}$}} & \multicolumn{1}{l|}{\multirow{2}{*}{$\Gamma$}} & \multicolumn{3}{c|}{Range of Fraction}                                                        \\ \cline{3-5} 
\multicolumn{1}{|l|}{}                                                            & \multicolumn{1}{l|}{}                          & \multicolumn{1}{c|}{Photosphere}     & \multicolumn{1}{c|}{Internal Shock}  & ICMART          \\ \hline
\multirow{3}{*}{10}                                                               & 100                                            & \multicolumn{1}{c|}{Rule out}        & \multicolumn{1}{c|}{Rule out}        & Rule out        \\ \cline{2-5} 
                                                                                  & 300                                            & \multicolumn{1}{c|}{{[}0, 6.1\%{]}} & \multicolumn{1}{c|}{{[}0, 8.2\%{]}} & {[}91.8\%, 1{]} \\ \cline{2-5} 
                                                                                  & 500                                            & \multicolumn{1}{c|}{{[}0, 25.5\%{]}}  & \multicolumn{1}{c|}{{[}0, 48.0\%{]}}   & {[}52.0\%, 1{]}    \\ \hline
3                                                                                 & \multirow{3}{*}{300}                           & \multicolumn{1}{c|}{No constraint} & \multicolumn{1}{c|}{No constraint}   & No constraint   \\ \cline{1-1} \cline{3-5} 
5                                                                                 &                                                & \multicolumn{1}{c|}{{[}0, 78.5\%{]}} & \multicolumn{1}{c|}{No constraint}   & No constraint   \\ \cline{1-1} \cline{3-5} 
10                                                                                &                                                & \multicolumn{1}{c|}{{[}0, 6.1\%{]}} & \multicolumn{1}{c|}{{[}0, 8.2\%{]}} & {[}91.8\%, 1{]} \\ \hline
\end{tabular}
\label{table2}
\end{table}

\end{document}